\documentclass[10pt, conference, letterpaper]{IEEEtran}

\usepackage{cite}
\usepackage{graphicx}
\usepackage{subcaption}
\usepackage{refstyle}
\usepackage[cmex10]{amsmath}
\usepackage{algorithmic}
\usepackage{array}
\usepackage{url}
\usepackage{fontenc}
\usepackage[utf8]{inputenc}
\usepackage{mathtools}
\usepackage{tabularx}
\usepackage{tabu}
\usepackage{float}
\usepackage{multirow}
\usepackage{siunitx}
\usepackage{algorithmic}
\usepackage{psfrag}
\usepackage{pgfplots}
\usepackage{xcolor}

\usepackage{comment}
\usepackage{wrapfig}
\usepackage{nopageno} 
\usepackage[utf8]{inputenc}
\usepackage{amsmath,amssymb}
\usepackage{makecell}
\usepackage{booktabs}
\usepackage{enumitem}
\usepackage[ruled,vlined]{algorithm2e}
\SetKwInput{KwInput}{Input}                
\SetKwInput{KwOutput}{Output}              

\newsavebox{\ieeealgbox}

\newcolumntype{M}[1]{>{\arraybackslash}m{#1}}
\newcolumntype{N}{@{}m{0pt}@{}}

\usepackage{proof}

\begin{document}

\title{Learning from Peers at the Wireless Edge}

\author{\IEEEauthorblockN{Shuvam Chakraborty\IEEEauthorrefmark{1}, Hesham Mohammed\IEEEauthorrefmark{1} and Dola Saha}
\IEEEauthorblockA{
Department of Electrical \& Computer Engineering \\
University at Albany, SUNY, Albany, NY 12222 USA\\
\{schakraborty, hhussien, dsaha\} @albany.edu\\
\IEEEauthorrefmark{1}co-First Authors}
}


\maketitle

\begin{abstract}

The last mile connection is dominated by wireless links where heterogeneous nodes share the limited and already crowded electromagnetic spectrum. Current contention based decentralized wireless access system is reactive in nature to mitigate the interference. In this paper, we propose to use neural networks to learn and predict spectrum availability in a collaborative manner such that its availability can be predicted with a high accuracy to maximize wireless access and minimize interference between simultaneous links.
Edge nodes have a wide range of sensing and computation capabilities, while often using different operator networks, who might be reluctant to share their models. Hence, we introduce a peer to peer Federated Learning model, where a local model is trained based on the sensing results of each node and shared among its peers to create a global model. The need for a base station or access point to act as centralized parameter server is replaced by empowering the edge nodes as aggregators of the local models and minimizing the communication overhead for model transmission. We generate wireless channel access data, which is used to train the local models. Simulation results for both local and global models show over 95\% accuracy in predicting channel opportunities in various network topology.

\end{abstract}
\section{Introduction}

Exponential increase~\cite{cisco} in data capacity requirement for emerging applications can only be sustained by efficient usage of electromagnetic spectrum by a variety of heterogeneous devices. Efforts have been made to open up new spectrum, while several unlicensed and semi-licensed models have been proposed for a shared usage. Although, multiple operators will prevail for licensed access, large swaths of frequencies will be available for unlicensed use for different protocols. We envision that future intelligent wireless networks will be able to make distributed decisions on wireless channel access without any aid from the centralized base station.

\begin{figure}
    \centering
    \includegraphics[width=0.75\linewidth]{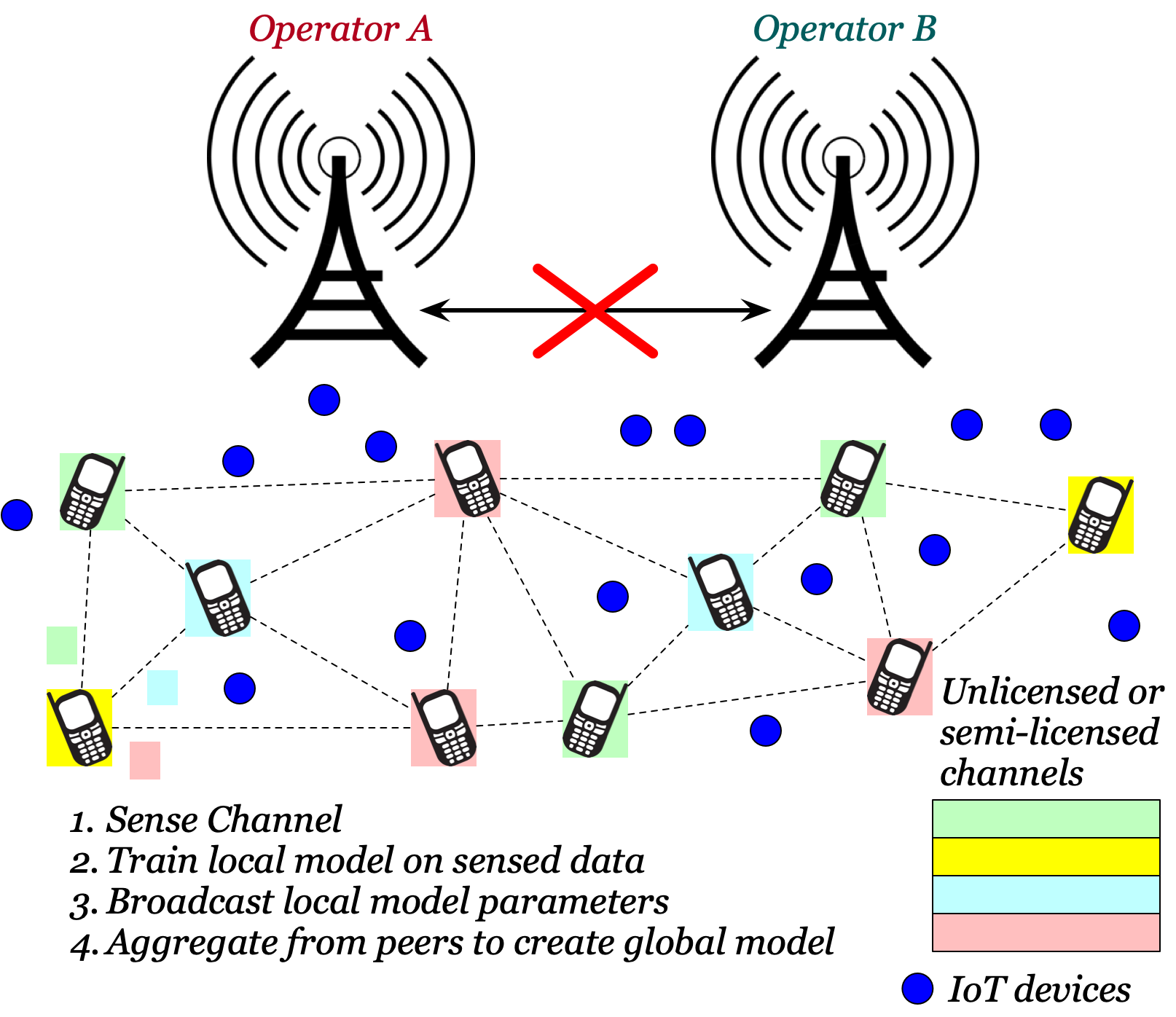}
    \caption{Future Intelligent Wireless Network.}
    \label{fig:system}
\end{figure}

Distributed wireless channel access is performed using carrier-sense and backoff mechanisms as in Wi-Fi~\cite{802_11_spec}.
As the system only reacts to collisions, much of the time is wasted in sensing, backoff and collisions as the number of nodes in the system increases~\cite{wi-fi_bianchi}. If an accurate collaborative prediction system is appointed, we will notice a better usage of the available spectrum.
Machine learning based wireless systems have received attraction in recent years to learn hidden parameters in a system, which are difficult to model. In this scenario, traditional machine learning approaches require centralizing the training data and inference processes on a single data center. Due to the propagation characteristics of radio frequencies, wireless channel is inherently distributed, and has to be measured and learned at each node for optimum performance. A base station's view of wireless channel could be completely different from a mobile terminal's view, specially when they are spatially separated. Hidden terminal problems cannot be solved by a centralized entity when multiple parties share the radio frequency spectrum.
At the same time, many channel properties may overlap, which can be similar in the vicinity and learned from neighbors. Sensing at the mobile terminals and sending the data to the base station is infeasible because of communication costs. Moreover, there are trust and privacy issues, which deters the operators to share their data. With these challenges, we design our protocol to predict wireless channel availability in a distributed wireless network.



Figure~\ref{fig:system} shows the last mile future network, which mainly will constitute 1) Base stations, operated by different operators providing Internet access to mobile terminals and IoT devices, 2) IoT devices, densely deployed and often with limited sensing and computation capabilities, 3) Mobile terminals, including smartphones, tablets etc., which are capable of sensing and transmitting in a wide variety of frequency bands. Also, they may not have the capability to monitor all the available channels all the time, but should be able to use any of the channels when the transmission opportunity exists. One of the major issues of decentralized wireless access is also the hidden terminal problem, which cannot be mitigated by sensing at the transmitter. This requires learning channel availability at the intended receiver (one hop neighbor) and cannot be a localized decision. 
This is precisely the reason where we deploy Federated Learning~\cite{fed_comm_efficiency_16, agnostic_fed} to predict the channel availability in each node. The first step of our system is to sense the channel, which the mobile terminals choose depending on any specified criteria. Based on the sensing data, it trains a local neural network model to predict the channel availability. Then, it broadcasts the trained local model parameters to its neighbors using a shared control channel. These neighbors can be connected to different operators, but can form an overlay network with peers to share learned models over a common unlicensed channel. Once a node receives local models from its neighbors, it 1) concatenates the models for which it does not have the data and 2) aggregates the model by averaging the model parameters from its neighbors. The first case helps a node to learn channel availability quickly from neighbor, which it has not sensed and thus does not have a local model. The second case addresses the hidden terminal problem by considering channel prediction models of it's one-hop neighbors.

Our protocol does not require a centralized parameter server, since we deploy global model aggregation at each node. Channel availability prediction does not require to propagate multiple hops as it depends on the interference that a transmitter can create at another intended receiver. Hence, the global model is also small enough to be implemented at the edge. It is to be noted here that current smartphones already deploy neural network models in GPUs or neural processors for efficient image processing. Hence, our assumption of deploying a local neural network model on these smartphones at the edge of the network is quite realistic.

\section{Related work}
Federated learning~\cite{fed_comm_efficiency_16} was proposed to increase communication efficiency where the entire data-set is not readily available to the central server and mobile nodes have a small fraction of that data available to them. They use the local data to learn the local model and share only the model parameters with parameter server (PS). The model parameters are aggregated in the centralized PS to generate the global model, which is shared with the mobile nodes. There has been numerous applications of federated learning to model various aspects of wireless systems, none of those have attempted to make the system completely decentralized removing the need of any parameter server. 
Authors in~\cite{hu2019decentralized} proposed a model segment level decentralized federated learning to pull the models from participating nodes. 
Authors have taken a segmented update approach in ~\cite{8815498, DBLP:journals/corr/abs-1807-02515}, which in spite of being a fully decentralized approach, needs number of nodes for each gossip segment to be precisely defined for most efficient model update. This is not needed in our system design. \cite{8737464} explores the effect of varying number of nodes updating simultaneously to the parameter server. A federated learning approach for packet classification has been discussed in~\cite{bakopoulou2019federated}, which also requires parameter server to aggregate the model.
~\cite{abad2019hierarchical, 232971, park2018wireless, yang2019scheduling, ahn2019wireless, zhu2018broadband, vu2019cellfree} incorporates various applications and model updates for Federated learning, which uses base stations as the parameter server. 
A peer to peer model of federated learning is proposed in~\cite{p2p_fedLearn_19} where the authors assumed the data is available to each mobile nodes. Also, there is an assumption that the data is fully orthogonal or uncorrelated. 
On the contrary, in our system, the neighbors in close proximity will have highly correlated data based on channel sensing. Hence, none of the above mentioned models or solutions can be applied directly in our system.



\section{Background}


In this section we describe the concept of federated learning and why it is so well suited for our problem of channel sensing and prediction. 
Federated Learning enables distributed devices to collaboratively learn a shared prediction model while keeping all the training data on the device. Once trained, the updated parameters are aggregated in a centralized parameter server to create a global model.
Assuming $n$ nodes are present in a network, and $\theta_{i}$ is the local model parameter matrix of the node $i$, then the aggregator creates a global model $\Theta$ as shown in equation~\ref{eq:global}.

\begin{equation}
\label{eq:global}
    \frac{1}{n} \sum_{i=1}^{n} \theta_{i} = \Theta
\end{equation}


Any wireless communication system is inherently a distributed system. Conventional ML systems work on the assumption of having the entire data-set and processing capability available in a central server. It is not feasible in our case not only due to privacy reasons, but due to high volume of data that needs to be shared for training purposes yielding high communication costs. Consequently, decentralized approach is a lucrative solution that incurs minimum communication overhead and computation costs. 

\section{Problem Formulation}

In this section we demonstrate peer to peer based federated learning system for wireless networks for predicting channel availability. 
The notion of distributed learning regime lies in two possible scenarios: data parallelisation and model parallelisation. While federated learning predominantly exploits the data parallelisation by using the same training model with orthogonal or non-overlapping data-set. In wireless systems, the data might overlap, thus providing higher priority for the overlapping data, as this is informed by frequent appearance of relative parameter. 
We assume $N$ number mobile nodes in the network, where $i^{th}$ node is denoted by $N_i$.
All the nodes are acting as wireless sensors denoted by set $\mathcal{N}$ and are participating in the distributed learning. 
Each node has their model generated from channel sensing results of that node itself, thus guaranteeing only a local view of the wireless system owing to limited visibility of the mobile nodes. Each local data-set is denoted by $X_i$ and is accompanied by a label set $Y_i$, $i \in N$. Following our assumption that local data-sets have overlap in wireless networks, $\bigcap^n_{i=0}X_i\neq \emptyset$, where $\emptyset$ denotes the empty set.


Each node $N_i$ generates a local parameter set $\theta_i$, where $i$ denotes the node identity. These parameters are shared among the neighboring mobile nodes. 
Only the model parameters from the local model are shared with other nodes in a broadcast signal, as it does not include any raw data from the primary node leveraging the inherent data preserving nature of federated learning.
We assume the local models implemented in the mobile nodes has access to the local likelihood functions that generate each local weight matrix or parameter matrix $\theta_i$.

Based on our assumptions, the global parameter generated at the node $N_i$ can be denoted as $\Theta_i$, where
\begin{equation*}
\| Y_i - X_i\theta_i \|= \eta_1  \quad \texttt{and} \quad
\|Y_i - X_i\Theta_i \| = \eta_2 
\end{equation*}
where $\eta_2\leqslant\eta_1$ and  $\|.\|$ is the $L_{2}$ norm. Since $\eta_2$ denotes the error rate in predicting channel availability while using updated global parameter, it should be equal or less than the error rate using local models $\eta_1$, because of limited channel information shared in local models.

\section{Channel availability prediction Protocol}
\label{protocol}

\begin{figure}
    \centering
    \includegraphics[width=0.8\linewidth]{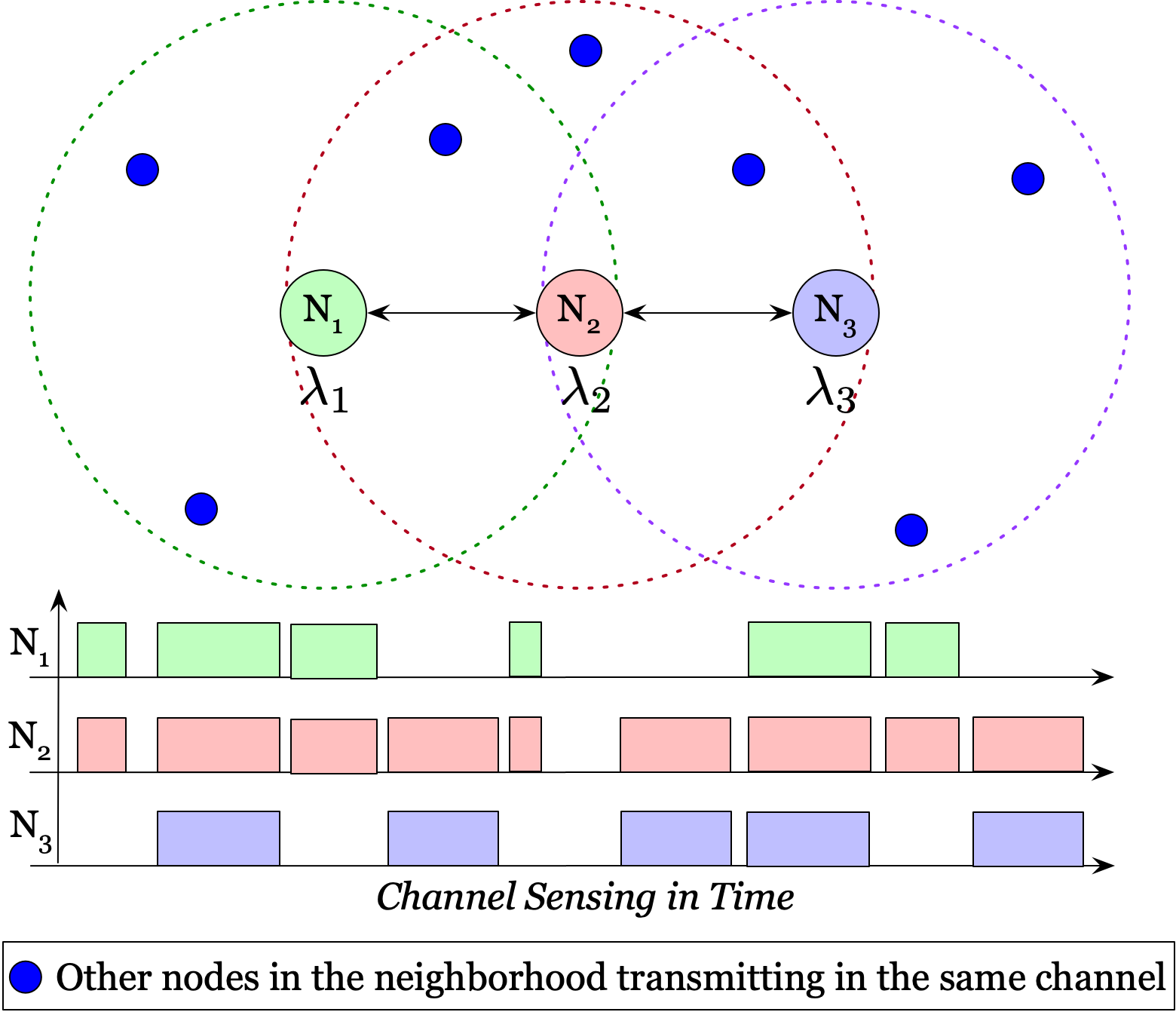}
    \caption{Channel Sensing and Local Model Exchange in Hidden Terminal Scenario.}
    \label{fig:hiddenTerm}
\end{figure}

In traditional CSMA-CA system, channel availability is sensed by a mobile node for a short duration and if it senses the channel busy, it backs-off for a duration $w$ randomly chosen from the contention window, which grows exponentially in every iteration if the node senses the channel to be busy. Furthermore low power mobile nodes can sense only one channel giving rise to uneven distribution of channel resource usage for each node. 
Figure~\ref{fig:hiddenTerm} shows a hidden terminal scenario, where nodes $N_1$ and $N_3$ are hidden to each other when they sense the channel and transmit at the same time to create interference at the receiver, node $N_2$. Hence, sensing locally and learning on only local sensing data will not address the hidden terminal issue. It is important to capture node $N_2$'s sensing information in the learning parameters of both $N_1$ and $N_3$. Thus, when local model $\theta_2$ of node $N_2$ gets propagated to both its neighbors, channel availability at receiver $N_2$ is also incorporated in the aggregated model. 
It is to be noted here that sensing the channel creates a prediction for all the transmissions near that node. Hence, even when IoT devices or other nodes are not sensing the channel, their transmission characteristics are captured by one-hop neighbors who are sensing the channel. 

\subsection{Channel sensing}

Multiple traffic arrival rates (multiple varying $\lambda$s) are incorporated in the channel traffic model, where individual traffic arrival follows Poisson distribution and different possible arrivals are uniformly distributed in time. 
These traffic flows may be generated by one node or multiple nodes, but, when transmitted in a channel, is sensed by all neighbors which are sensing or receiving in that channel.
For example, multiple IoT devices may generate different traffic rates, and might not be sensing the channel due to power constraints. However, a mobile node, if participating in sensing and collaboration, will sense the channel and observe it to be busy during the transmission period. For example, if multiple nodes around node $N_1$ in figure~\ref{fig:hiddenTerm} generates traffic at different rates and transmits them, then sensing at $N_1$ will capture all those times as channel being busy. Thus, $\lambda_1$ is a combination of multiple traffic patterns. Also, there are multiple nodes that are common in one-hop neighborhood, thus there is a significant overlap of data among one-hop neighbors.
The mobile nodes sense the channel for a small time period $\delta$, where $\delta \ll L_{pkt}$, and $L_{pkt}$ is the minimum packet transmission duration in the network.
In each $\delta$, if the mobile terminal senses the channel busy for any duration, it indicates the channel to be busy (denoted as 1) for that time, thus discretizing the channel output and generating a sequence of bits that encodes channel sensing result as a binary time-series.

\subsection{Training Local Model}
Each node $N_i$ generates the time-series as channel sensing result, which includes channel activity sensed within the coverage area of this node. This time-series is mapped is mapped into one-hot code and fed to a two layer LSTM network, which generates the local parameter set $\theta_i$ for channel prediction depending on sensing data from node $N_i$ only.

\subsection{Model Sharing}
Every node shares their locally learned parameter matrix $\theta_i$, for node $i$ as a broadcast packet, thus sharing its local model only to all one hop neighbors. These local models contain parameters learned from only the local limited view of the source node. The parameters learned from the neighbors are not shared, thus limiting the model propagation to one hop only. 


\subsection{Global Model Generation}

All nodes $N_j, j \neq i, j \in (1, M)$ are sensing the same channel thus seeing a part of the same network traffic as the primary node $N_i$ denoted by $\lambda_i$ along with other network traffic denoted by {$\lambda_1, \lambda_2,..., \lambda_M$}, 
where each of them are association of set of different arrival rates. In the figure~\ref{fig:hiddenTerm} we can only see three nodes, but we will generalize our discussion here. In this scenario all local parameter set $\theta_l$ generated from $N_l, l \in M$ will carry the information of network traffic $\lambda_i$. Thus  averaging, of the parameters will generate very high accuracy of the global model of node $N_i$. Averaging will reduce weights of network traffic contributed by $\lambda_k$s. Thus setting $w_i = 1$ generates 98\% accuracy for channel prediction of node $N_i$
So the global model becomes,
\begin{equation}
   \theta_i + \frac{1}{M} \sum_{l=1}^{M} w_l(\theta_{l}+\eta_{l}) = \Theta_i
  \end{equation}
 where $w_l$=1, $l \in M$

\subsection{Learning parameters for orthogonal channels}

There might appear another scenario where nodes sense different channels even in a similar network as shown in Figure~\ref{fig:hiddenTerm} consequently generating local model parameters that are entirely uncorrelated, thus aggregating parameters following any algebraic operation is not feasible. So to predict other channel availability we have to store the model parameters and generate a concatenated global model for channel prediction. Here as we are storing different models and since it is a multi-node update process, there needs to be an optimum number of shared models to be stored while dealing with memory constrained edge devices.

\section{Neural Network Architecture}

\begin{figure}
    \centering
\begin{subfigure}[b]{0.27\textwidth}
    \includegraphics[width=0.9\linewidth]{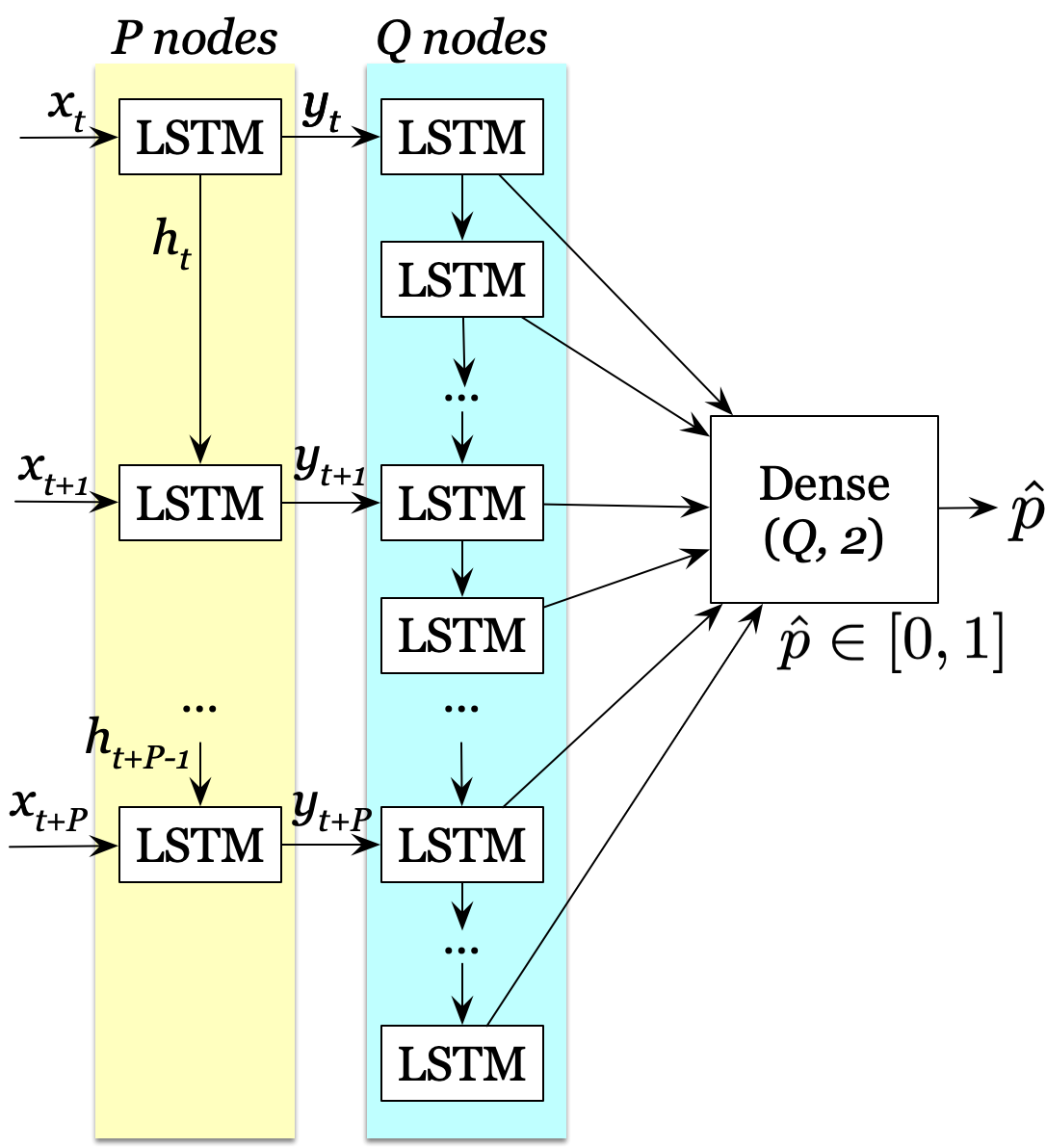}
    \caption{Neural Network Architecture.}
    \label{fig:nn}
\end{subfigure}
\begin{subfigure}[b]{0.2\textwidth}
    \includegraphics[width=0.9\linewidth]{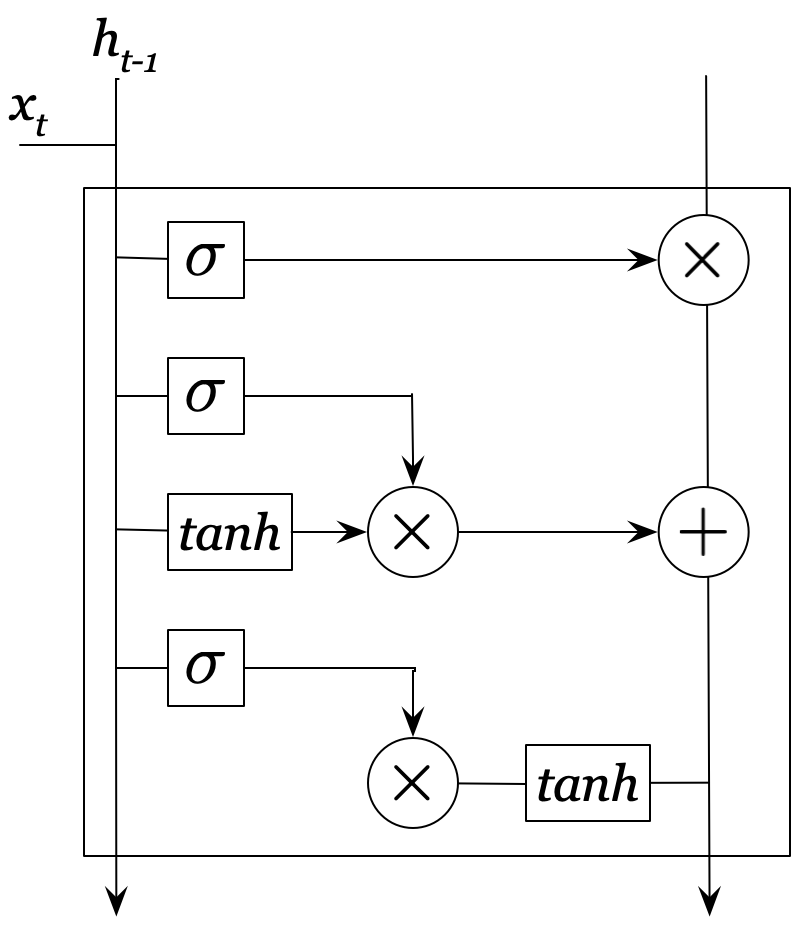}
    \caption{LSTM cell.}
    \label{fig:LSTM}
\end{subfigure}
\caption{Neural Network architecture in each edge node.}
\label{fig:nn_edge}
\end{figure}

Traditional neural networks (NN) (i.e., feed-forward networks and CNN) are not capable of learning data sequences such as text prediction since the output of the traditional NN depends on the current input and is given by:
\begin{equation}
    Y_{i}= f(w,b,X_{i})
    \label{eq:NN output}
\end{equation}
where $f$ is the NN activation function, $w$ and $b$ are the weights and biases of the NN, and $X_{i}$ and $Y_{i}$ are the input and the output respectively.

Recurrent NN (RNN)~\cite{book} solves this problem by making the output of the RNN depends on the current input and the input state. The input state relies on the history of the previous inputs and the outputs of the RNN. The output of the RNN is given by :
\begin{equation}
    (Y_{t},h_{t})= f(w,b,X_{t},h_{t-1})
    \label{eq:RNN output}
\end{equation}
where $h_{t}$ is the state of the RNN at time $t$.
However the RNN can not capture the long term dependencies, since the output depends on only one cell state ash shown in (\ref{eq:RNN output}). Long Short Term Memory networks (LSTMs)~\cite{LSTM1} are a special kind of RNNs, which are capable of capturing the sequence dependencies both long and short term. 

\subsection{Neural Network Structure}

We use LSTMs in the local model at each edge node, the structure of which is shown in figure~\ref{fig:nn_edge}. 
As depicted in figure~\ref{fig:nn} the LSTM network has 2 layers including $P$ nodes in the first layer and $Q$ nodes in the second layer followed by a dense layer that generates the trained parameters. We have chosen a two layer LSTM network along with a dense layer before output, with neurons \{$P$,$Q$\} pairs, where outputs from P neurons of input layer is mapped in an one to many fashion to neurons of second layer with $Q$ elements, which ultimately generates a parameter set of dimension ($Q$,2).

Each cell of LSTM network is based on two main components. The first part is the conveyor belt, which shares the network history among all the LSTM cells. The second part is the component of the LSTM cell. The LSTM cell consists of three sigmoid functions, which act as three main gates as shown in figure~\ref{fig:LSTM}. The first gate allows the cell input update the conveyor belt. The second one decides whether the cell state is affected by both the conveyor belt (i.e the Network history) and the cell input or depends on the conveyor belt. The last one controls the cell output such that the cell output depends on both the input and the cell state or results from the cell state only.

\subsection{Data mapping}
In this work, we consider the channel state prediction which has two states, either idle or busy channel. This representation is not suitable for neural networks (NN) since they deal with the real numbers. So, we transform the data from binary representation to a supported data representation to construct the data set used to train the LSTM.

Let $m$ be the set of all possible events. We use one-of-$m$ representations for the channel state, given by  

\begin{equation}
    p_{k}= [1(x_{k} = Z_{1}),1(x_{k} = Z_{2}),1(x_{k} = Z_{3}),..,1(x_{k} = Z_{m})]^{T}
\end{equation}
Therefore, the element corresponding to the event equals to 1 while the others are 0 (i.e one-hot encoding). In this work, we only consider one channel measurement at a time (i.e $m=2$). However this representation is also valid if the node can sense more than one channel at a single time slot. Note that $P_{k}$ can be considered as the probability mass function (PMF) of the event to happen.

The output layer of the LSTM is a softmax layer. The softmax layer output $\hat{p_{k}}\in [0,1]$ is the probability vector of the transmitted message. The softmax function is given by:
\begin{equation}
    \label{eq:softmax}
    \sigma(x_{j})=\frac{e^{x_{j}}}{\sum_{i}e^{x_{i}}}
\end{equation}
where $x_{j}$ is the component $j$ in the vector $x$.

In the training phase, The LSTM updates its parameters in each training epoch to achieve the optimal parameter
\begin{equation}
    w^{*} = \min L(\hat{p}=p/p)
    \label{eq:opt}
\end{equation}
where $L$ is the loss function between the predicted and the actual states, which is similar to maximize log likelihood that acts as the best estimator. 

\subsection{Loss Function}
The Loss function is used to adjust the weights and biases to map the LSTM prediction to the actual targets included in the training set. The optimization problem in (\ref{eq:opt}) is solved by applying stochastic gradient descent (SGD) using a cross-entropy loss function which is given by:
\begin{equation}
    L_{cross} = H(\hat{p},p)= H(p)+ D_{KL}(p||\hat{p})
    \label{eq:crossentropy}
\end{equation}
where $H$ is the entropy, and $D_{KL}(||)$ is the Kullback-
Leibler divergence \cite{Cross}. Note that, minimizing $L_{cross}$ or $D_{KL}(||)$ is equivalent to maximizing the likelihood between the predict probability and the actual occurrence of an event.

\section{Experiment and Evaluation}
\label{sec:evaluation}
\subsection{Experimentation Setup}

We have generated the network topology in MATLAB assuming varying Poisson arrival rates with different packet sizes.  For experimentation, we choose two different set-ups.The first set of experiment includes one primary node and three different neighbors, sensing a part of common network traffic denoted by average arrival rate $\lambda_i$ that is sensed by primary node $N_i$ as well, where as in the second set there are five neighbors introducing further variation to channel traffic. Motivation behind choosing such a setup was to demonstrate the local model and global model update of that primary node for any real network scenario with varying number of neighbors, where there will be some overlap of sensing data among the neighbors. 
The neighbouring nodes sense additional network traffic on the same channel denoted by $\lambda_1$, $\lambda_2$, etc. where each of these terms are associated with multiple different packet arrival rates. We have tested this case for five neighbouring nodes as well, incurring further variance and consequently higher mutual information between the nodes to learn.


We have chosen the maximum packet size to be equal to standard wireless TCP packet size 2312 bytes along with total 52 bytes of MAC and IP headers. The packet length varies between 2000 to 2364 bytes to emulate a real network traffic. Any edge sensor senses the channel for a duration of 20$\mu s$, which is one of parameters of the DIFS (DCF interframe spacing) times in IEEE 802.11n standard. The sensing data is generated as a time-series from a seed with channel traffic distributed as Poisson distribution and effectively the channel idle time as an exponential distribution with labels 0 (denoting channel idle) and 1 (denoting channel busy) for a total channel sensing duration of \textit{5 seconds}. All our experiments require only 5 seconds worth of data for the local model to be trained. No training is required after aggregation, essentially making it practical for deployment. This generates $250,000$ instances of data in the time-series for training the LSTM network, 10\% of training data-set, generated using a different seed is used for validation and the entire training is implemented on an Intel NUC (NUC7i7BNH) with i7-7567U processor and 16GB DDR4 memory, without using any acceleration units. 
The different packet arrival rates used for experimental set-up has been shown in table~\ref{tab:params} for all different experimental setup.
According to the figure~\ref{fig:nn} we have chosen two different sets of \{$P$, $Q$\} pairs, \{60, 120\} and \{5, 5\}, which will be denoted in the following evaluation section as $T_b$ and $T_s$. 

\begin{table}
\caption{Packet arrival rates in different network topology}
\vspace{-10pt}
\label{tab:params}
\begin{center}
\begin{tabular}{ |l|l|l| } 
\toprule
 \makecell[l]{Network \\Topology} & 
 \makecell[l]{Arrival Rate\\ of Primary Node} &
 \makecell[l]{Arrival Rate\\ of Neighbors} \\
\toprule
 \makecell[l]{3 neighbors}
  & $\lambda_{i} = 5.0$ &  \makecell[l]{$\lambda_{1} = \{5.0, 9.5, 12.0\}$\\ $\lambda_{2} = \{5.0, 8.6, 10.5\}$ \\ $\lambda_{3} = \{5.0, 16.0, 6.0\}$} \\ 
 \hline
\makecell[l]{5 neighbors} &  $\lambda_{i} = 5.0$ & \makecell[l]{$\lambda_{1} = \{5.0, 9.5, 12.0\}$\\ $\lambda_{2} = \{5.0, 8.6, 10.5\}$ \\ $\lambda_{3} = \{5.0, 16.0, 6.0\}$ \\ $\lambda_{4} = \{5.0, 15.8, 21.0\}$\\ $\lambda_{5} = \{5.0, 2.8, 13.0\}$} \\
\toprule
\end{tabular}
\end{center}
\end{table}

\subsection{Evaluation}

\subsubsection{Performance of Local Model}
\begin{figure}
    \centering
    \begin{subfigure}[b]{0.24\textwidth}
    \includegraphics[width=\textwidth]{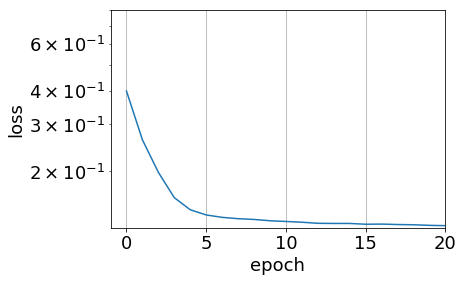}
    \caption{Local model loss function.}
    \label{fig:loss}
\end{subfigure}
\begin{subfigure}[b]{0.24\textwidth}
    \includegraphics[width=\textwidth]{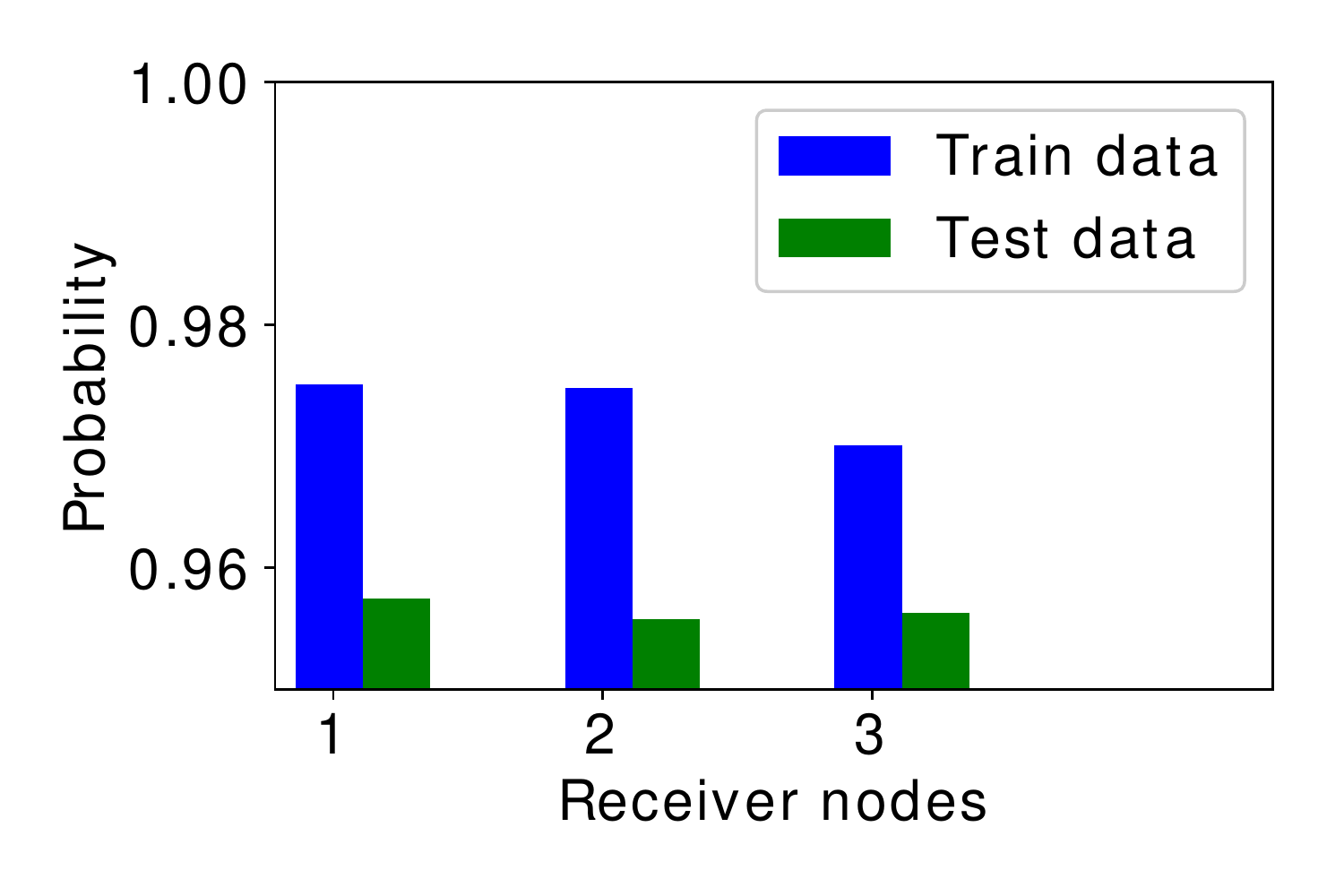}
    \caption{Prediction accuracy.}
    \label{fig:prob}
\end{subfigure}
\caption{Performance of local model (Node 1).}
\end{figure}

Figure ~\ref{fig:loss} shows the loss curve during the training of a local model for edge node 1. The other nodes' loss curves for local models overlap with it and hence is not shown in the graph. During training, network $T_b$ reaches about 97.8\% training accuracy in 20 epochs, which equals to approximate computation time of 250 seconds. It achieves 96\% validation accuracy in predicting channel occupancy only with local model and this value stays same as demonstrated in figure~\ref{fig:prob} for three different nodes with their local models.

\subsubsection{Performance of Global Model}
\begin{figure}
    \centering
    \begin{subfigure}[b]{0.49\linewidth}
        \includegraphics[width=\linewidth]{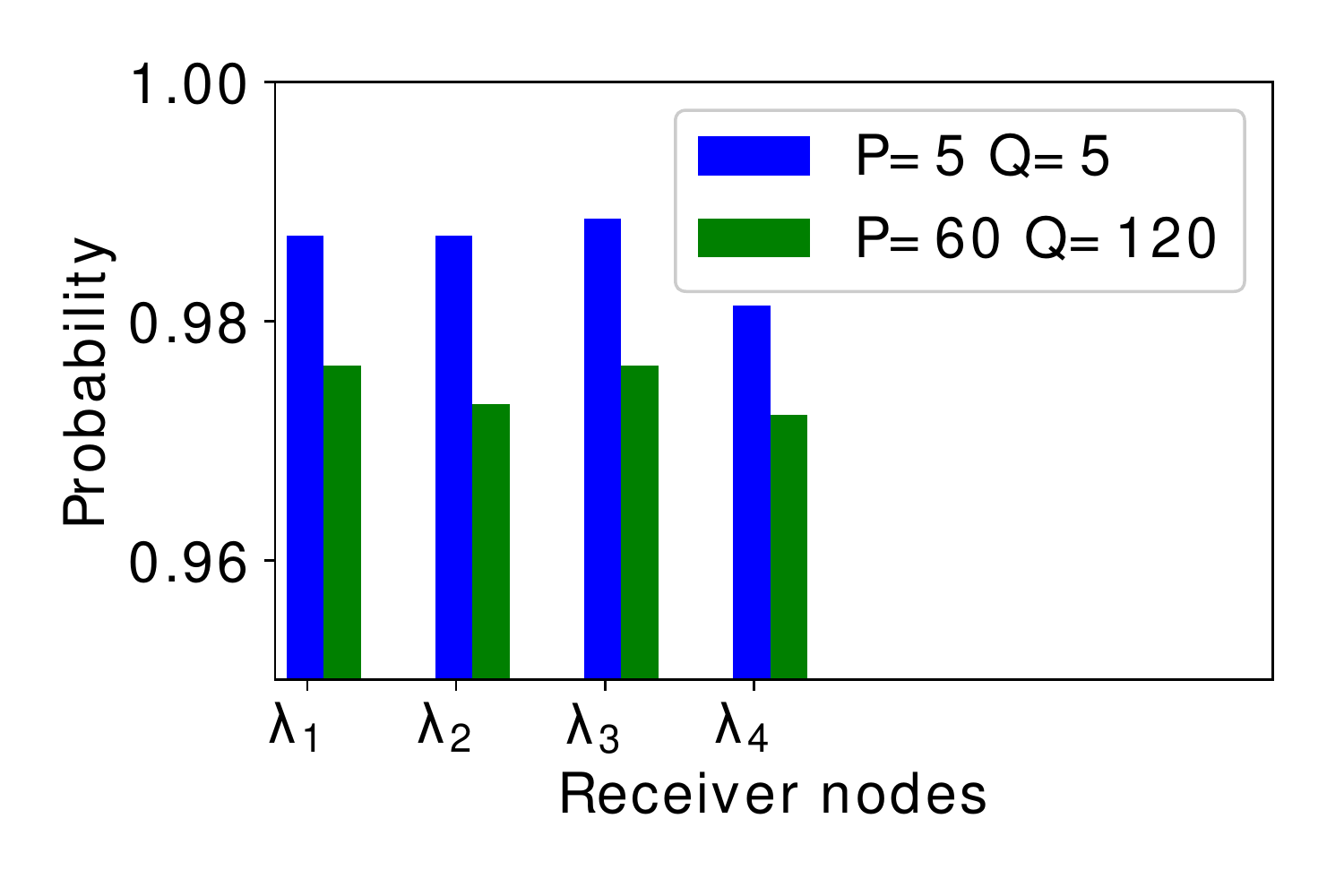}
        \caption{Three Nodes}
        \label{fig:prob_Av}
    \end{subfigure}
    \begin{subfigure}[b]{0.49\linewidth}
        \includegraphics[width=\linewidth]{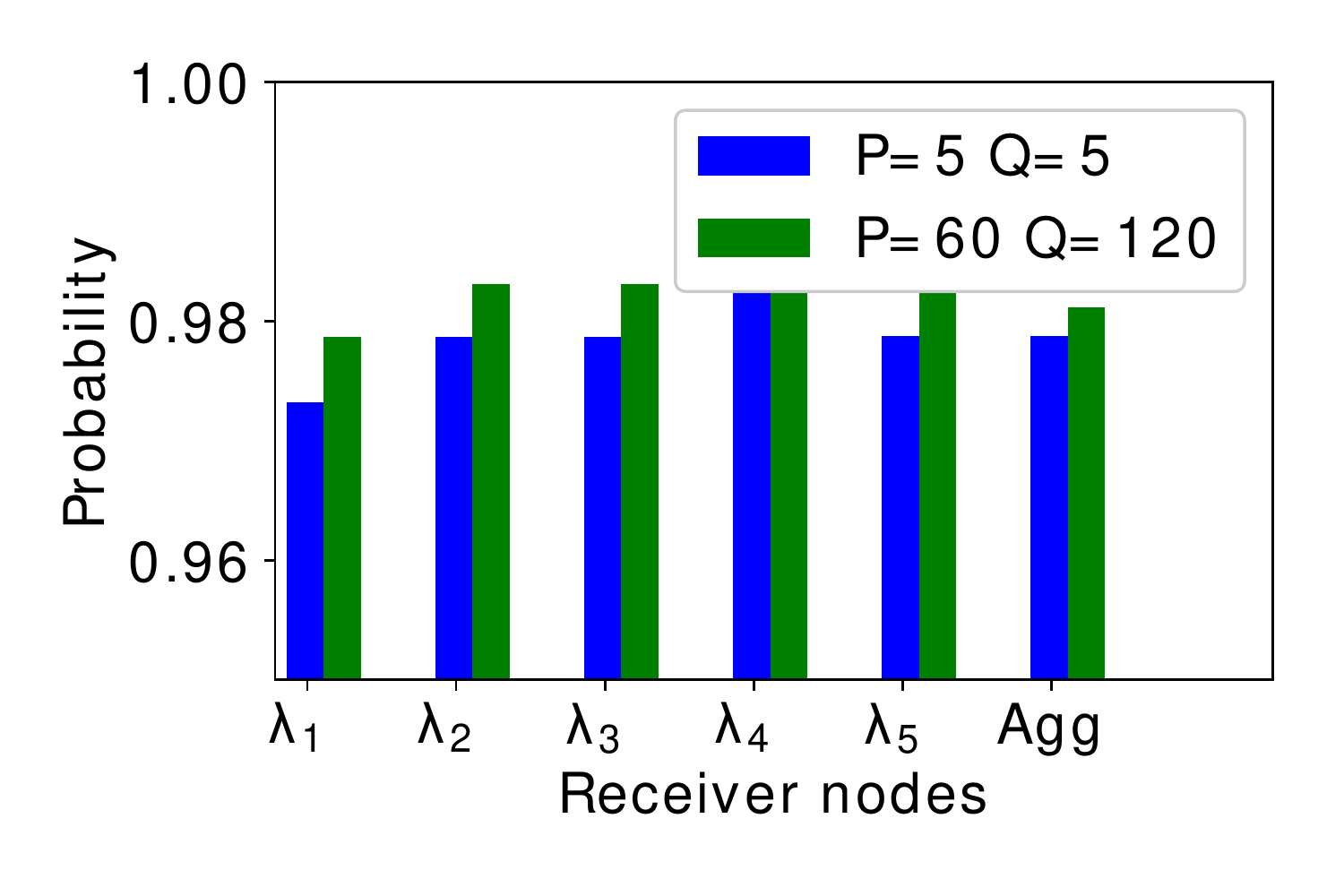}
        \caption{Five nodes.}
        \label{fig:prob_Av_5}
    \end{subfigure}
    \caption{Performance of global model for multiple nodes and different size model.}
    \label{fig:loss_cmp}
\end{figure}

We have tested accuracy of global models with three neighbors and five neighbors. Figure~\ref{fig:prob_Av} shows the channel prediction accuracy for local models of each neighbors as well as that of the aggregated global model of Node 1 using both, network $T_b$ and $T_s$. 
Leveraging accuracy of LSTM networks in prediction of sequential data, both the networks $T_b$ and $T_s$ are able to achieve about 98\% validation accuracy. 
$T_b$ required 40 epochs to generate weights that helps global model to predict the channel with an accuracy of 98.12\%, which is equivalent to computation time of 250 seconds.
With the same input data, $T_s$ required 400 epochs, though total number of up-gradable neuron weights being much less it incurred a computation time of 240 seconds, which is comparable to $T_b$. We intend to use the smaller neural network, $T_s$, which achieves similar accuracy as the larger network, $T_b$, but requires much smaller footprint to be implemented in hardware and a smaller model update packet to be transmitted over the air to the neighbors, thus reducing communication overhead. 
These times reported in seconds do not use any acceleration units, like Neural Processing Units (NPUs) and Graphics Processing Units (GPUs), which are prevalent in current smartphones. In other words, when deployed in edge nodes, the computation time will be even less, thus making it a practical choice for channel availability prediction.

\begin{figure}
    \centering
    \begin{subfigure}[b]{0.48\linewidth}
        \includegraphics[width=\textwidth]{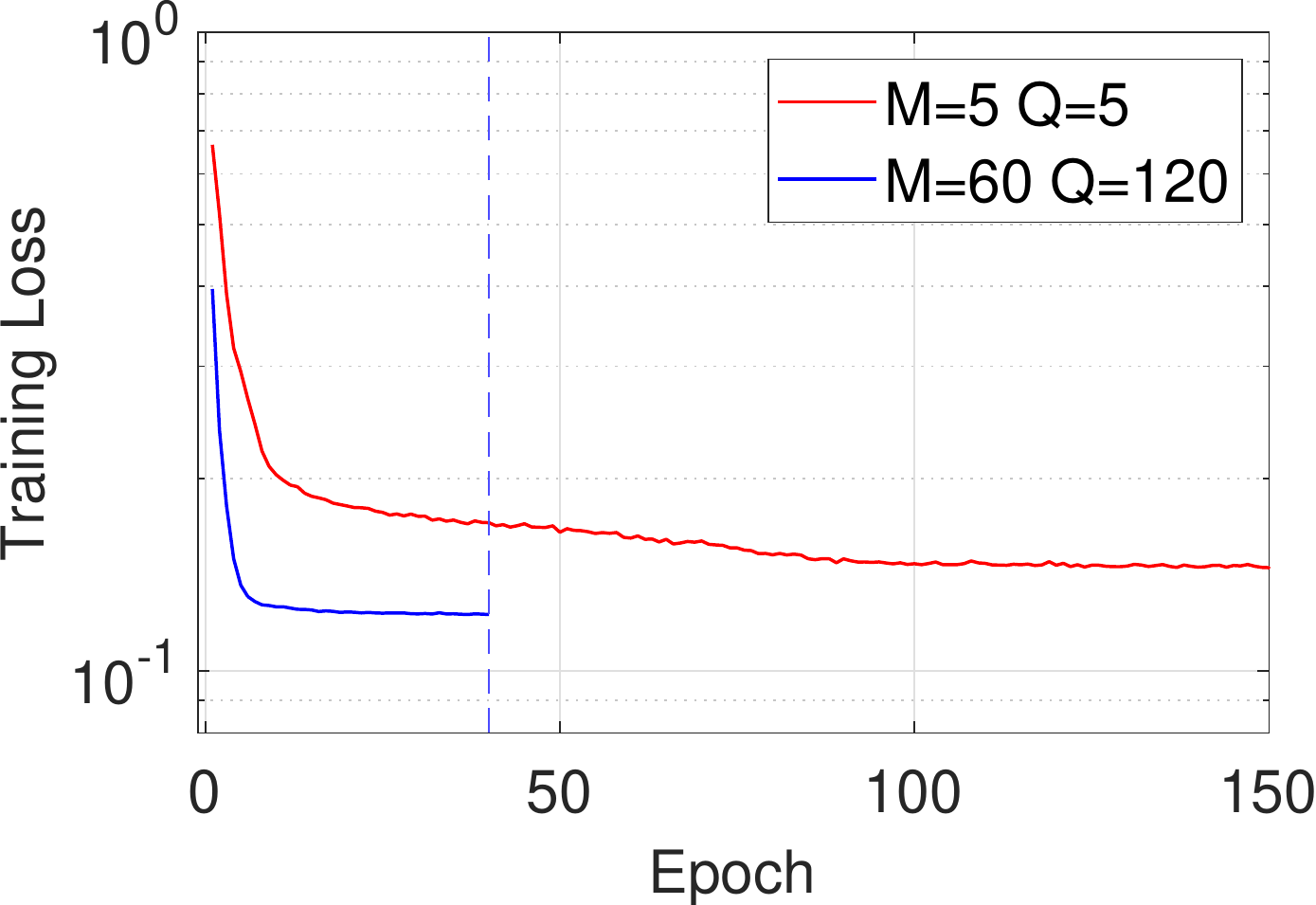}
        \caption{Loss function with increasing number of epochs.}
        \label{fig:loss_e}
    \end{subfigure}
    ~ 
    \begin{subfigure}[b]{0.48\linewidth}
        \includegraphics[width=\textwidth]{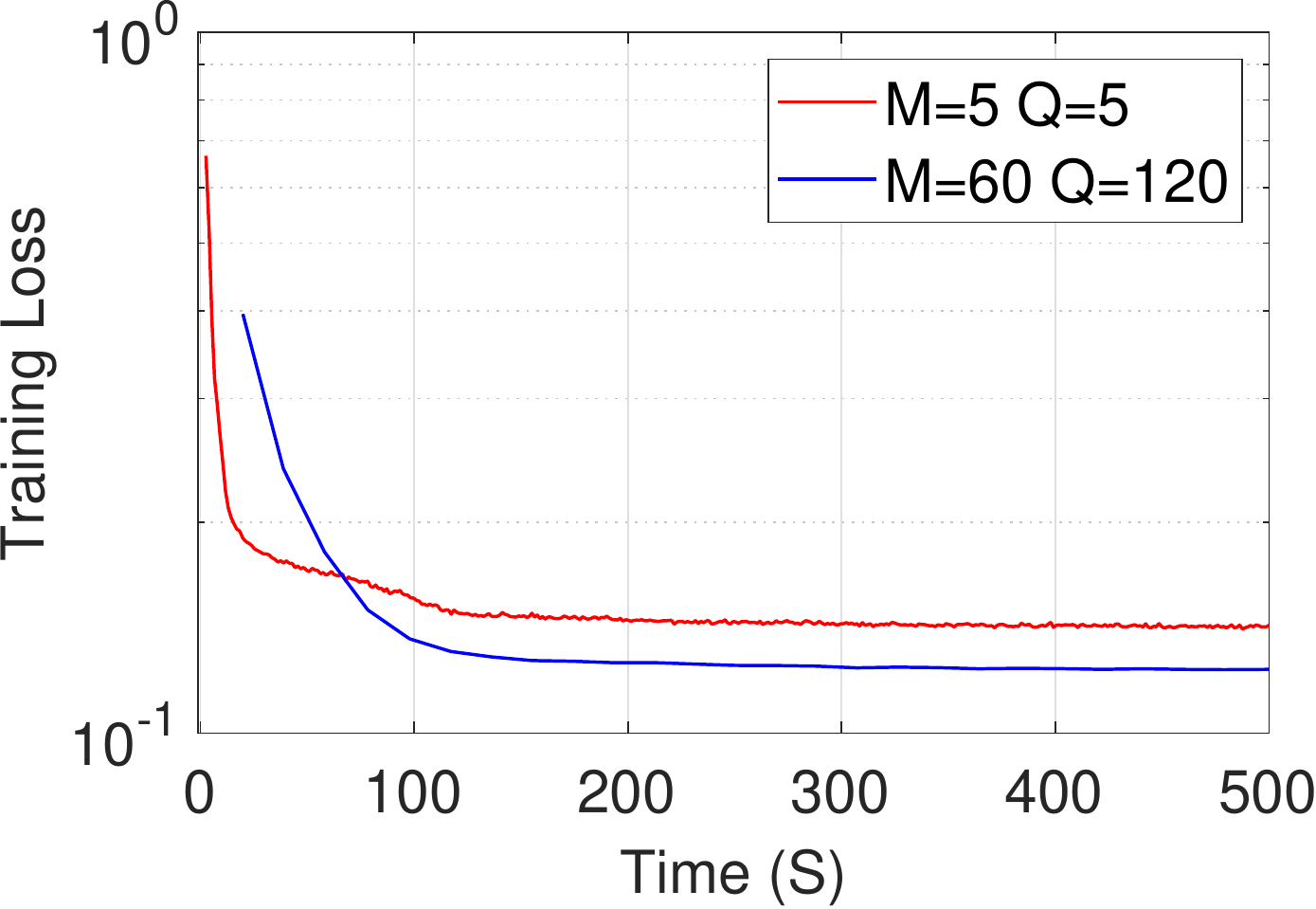}
        \caption{Loss function with computation time.}
        \label{fig:loss_t}
    \end{subfigure}
    \caption{Loss function of two different model sizes for the same node (Node 1).}
    \label{fig:loss_cmp}
\end{figure}

\subsubsection{Effect of Size of The Neural Network}
\label{sec:model_size}
The two LSTM networks used $T_b$ and $T_s$ have similar structure, thus performing in the same way, but $T_s$ is preferrable for the following properties:
\begin{itemize}[leftmargin=*]
    \item $T_s$ has 392 parameters (in floating point) in local model yielding 1512 bytes, which needs to be transmitted over the air to share with neighbors. On the other hand, $T_b$ has 102,252 parameters in local model yielding a size of 0.4 Megabytes. Since the local model has to be shared among peers, $T_b$ incurs higher computation and communication overhead than $T_s$.
    \item As shown in figure~\ref{fig:loss_e}, $T_s$ requires 400 epochs to reach the similar accuracy as of $T_b$, which it achieves in 40 epochs. But if we notice the computation time requirement, both networks converge at the same time as shown in figure~\ref{fig:loss_cmp}, thus reaching equivalent accuracy in same computation time.
    \item Since our primary implementation is for edge nodes with power and hardware resource constraints, the smaller network is a natural preference due to lower footprint and computation requirement.
    \end{itemize}
Thus even though in higher variance channel traffic $T_b$ generates slightly better results there are better trade-offs to opt for $T_s$ for deploying in edge nodes.

\section{Conclusion and future work}
In this paper, we have proposed a distributed framework for peer-to-peer based federated learning to reduce the need for centralized parameter server. Our evaluation shows it is highly effective in predicting channel availability in a wireless ad hoc network. Future work will require evaluation of the system in a larger network with a variety of channel access mechanism. Future exploration may include edge nodes to be able to sense disjoint channel properties and aggregate them for transfer learning.



\bibliographystyle{IEEEtran}
\bibliography{ref.bib}
\end{document}